\documentclass[twocolumn,amsmath,amssymb,nofootinbib,floatfix]{revtex4}

\usepackage[pagewise]{lineno}
\usepackage{graphicx}
\usepackage{dcolumn}
\usepackage{bm}
\usepackage{hyperref}
\usepackage{comment}
\usepackage{color}
\usepackage{subfigure}
\usepackage[papersize={10.5in,12in}]{geometry}
\usepackage{appendix}

\makeatletter

\newcommand{\Rmnum}[1]{\expandafter\@slowromancap\romannumeral #1@}
\makeatother

\def\be{\begin{equation}}
	\def\ee{\end{equation}}
\def\bea{\begin{eqnarray}}
	\def\eea{\end{eqnarray}}

\begin{document}
\title{Classical Analysis of Non-Coherent Dark Matter to Photon Conversion in a Resonant Cavity}
\author{Puxian Wei}
\author{Ruifeng Zheng}
\author{Qiaoli Yang$^*$}

\affiliation{Department of Physics, College of Physics and Optoelectronic Engineering, Jinan University, Guangzhou 510632, China}	
\begin{abstract}
Both axion and dark photon dark matter are among the most promising candidates of dark matter. What we know with some confidence is that they exhibit a small velocity distribution $\delta v\lesssim v\sim 10^{-3}$c. In addition, their mass is small, resulting in a long de Broglie wavelength and a high particle number density. Their phase space distribution contains many uncertainties, so they could give rise to either a coherent or noncoherent wave on the laboratory scale. In this paper, we demonstrated that a resonant cavity can enhance noncoherent axion-to-photon or dark photon-to-photon transitions, and the resulting power is the same as in the coherence case. The classical picture explanation is that a cavity can resonant with multiple different sources simultaneously. This aligns with the quantum perspective, where the cavity boosts dark matter particles transitioning into photons similarly to the Purcell effect. This effect increases the density of states near resonance, regardless of the coherence nature of dark matter. Certainly, the induced microwave signals in a cavity are also non-coherent, and in such case, a single-photon readout may be required.
\end{abstract}

\maketitle

\def\thefootnote{*}\footnotetext{qiaoliyang@jnu.edu.cn}\def\thefootnote{\arabic{footnote}}

\section{Introduction}
\label{sec1}

Dark matter is one of the most fascinating puzzles in modern physics. Cosmological and astrophysical observations indicate that it constitutes around 27\% of the total energy density of the universe. In addition, dark matter is expected to be cold, with a non-relativistic speed on the order $v \sim 10^{-3}c$ \cite{Herzog-Arbeitman_2018}. Axions and dark photon dark matter are among the most promising candidates. Both are thought to be light, with masses $m \lesssim 10^{-4}$ eV \cite{MARSH20161,Jaeckel:2010ni}. Furthermore, their coupling to Standard Model particles is extremely weak. One of the effective methods to detect them involves the use of a resonant cavity with a high quality factor $Q$ \cite{PhysRevLett.51.1415,PhysRevD.32.2988}.

Axion and dark photon waves have a large de Broglie wavelength $\lambda \approx 10$ meters for $m \approx 10^{-5}$ eV. The spatial correlation length could degrade to some extent due to uncertainties in the wave vector directions, but it can generally be treated as spatially constant on the laboratory scale. Temporal decoherence, however, poses a greater challenge. Bosonic dark matter can be created from non-thermal mechanisms such as \cite{Preskill:1982cy,Abbott:1982af,Dine:1982ah,Nelson:2011sf}. However, the cosmic structure formation may result in
coherence or non-coherent dark matter halos, and there are still many uncertainties. Assuming that each particle has a random phase $\phi_i$, the wave function could oscillate completely randomly: $D = D_0 e^{i[\omega_0 t + \phi(t)]}$, where $\phi(t)$ is a random function. The estimation of coherent time $\tau\approx 1/\delta \nu$ is valid when the waves have fixed relative phases. However, even if the waves are monofrequency $\delta \nu=0$, but composed of random phases, their coherent time is finite.   

In this paper, we find that the classical picture allows the signal from noncoherent waves to be boosted by the quality factor $Q$. This is consistent with the quantum perspective \cite{Zheng:2024kxn}, where the cavity boost is similar to the Purcell effect, resulting in a higher density of states near resonance compared to free space \cite{Yang:2022uil}. Certainly, the resulting signal is not coherent in nature if dark matter wave is noncoherent and may therefore require a single-photon detector for readout.

In the following discussion, we use the standard classical derivation \cite{Krauss:1985ub}, with necessary modifications to accommodate the noncoherent wave case.

\section{axion to photon transition in a resonant cavity}
\label{sec6}
The single axion particle wave function can be written as:
\begin{equation}
    \begin{aligned}
a_i(\vec x,t)=A_0 \cos \left[m\left(1+\frac{v_i^2}{2}\right) t+m\vec v_i\cdot \vec x+\phi _{i} \right]~,
   \end{aligned}
 \end{equation}
where \( v_i \) is small (e.g., \( 10^{-3} \)c). The multiparticle wave function can become decoherent. Spatial decoherence arises primarily from misalignment of \( \vec{v}_i \). Due to the large de Broglie wavelength \( \lambda \) compared to laboratory scales, the spatial component of the wave function is generally negligible if the dark matter particles have a small mass. Temporal decoherence is driven mainly by \( \phi_i \). If $\phi_i$ is completely random, The dark matter field can be written as \cite{Foster:2017hbq}:
\begin{equation}\label{EQ1}
    \begin{aligned}
a(t)=&\frac{\sqrt{\rho_{DM} }}{m} \sum_{j}^{} \alpha _j\sqrt{f(v_j)\Delta v} \\
&\cdot \cos \left [m\left ( 1+\frac{v_{j}^{2} }{2}  \right )t+\phi _{j}    \right ] ~,
    \end{aligned}
\end{equation}
where \( f(v) \) is the dark matter particle velocity distribution, \( \rho_{\text{DM}} \) is the local dark matter density, and \( \alpha_j \) is a random variable following a Rayleigh distribution $P(\alpha_j)=\alpha_je^{-\alpha_j^2}$.

The Lagrangian for the coupled axion and Standard Model photon field is:
\begin{equation}
    \begin{aligned}
 \mathcal{L} _{ap}=&-\frac{1}{4}  F_{\mu \nu }F^{\mu \nu }+\frac{1}{2} \partial _\mu a\partial ^\mu a-g_{a\gamma \gamma }a\mathbf{E} \cdot \mathbf{B}-\frac{1}{2} m^2a^2~,
  \end{aligned}
 \end{equation}
where \( F_{\mu \nu} = \partial_{\mu} A_{\nu} - \partial_{\nu} A_{\mu} \) is the field strength, \( A_{\mu} \) is the photon field, and \( g_{a\gamma\gamma} \) is the axion-photon coupling constant. In addition to the axion interaction, the electromagnetic field in the cavity also experiences a damping effect due to energy dissipation. Near a particular resonance frequency \( \omega_n \), the damping coefficient can be written as \( \omega_n / Q_n \), where \( Q_n \) is the cavity quality factor at frequency $\omega_n$. The quality factor measures the energy loss of the cavity. There are theoretically infinite resonant modes, but in practice, only a few low-frequency modes may be useful. For example, most axion search experiments use the TM\(_{010}\) mode, which has a higher \( Q_n \) and larger electric field overlap with the magnetic field. Assuming the applied magnetic field is $B_0\vec z$, the equation of motion for the electromagnetic wave inside the cavity is
\begin{equation}\label{eq33}
    \begin{aligned}
 \nabla ^2\mathbf{E} -\partial _t^2\mathbf{E}-{\omega_n\over Q_n}\partial_{t}\mathbf{E} =-g_{a\gamma \gamma }B_0\partial _t^2a\vec{\mathbf{z} } ~.
   \end{aligned}
 \end{equation}
 The electric field can be expanded by the cavity modes \( \mathbf{e}_n(\mathbf{x}) \) as
\begin{equation}
    \begin{aligned}
\mathbf{E}(\mathbf{x} ,t)=\sum_{n}^{} E_n(t)\mathbf{e} _n(\mathbf{x} )~
   \end{aligned}
 \end{equation}
where $\mathbf{e}_n$ satisfy the boundary condition of the cavity and the equation of motion:
 \begin{equation}
    \begin{aligned}
(\omega _n^2+\nabla ^2)\mathbf{e} _n(\mathbf{x} )=0~.
  \end{aligned}
 \end{equation}
By performing a Fourier transform, Eq.(\ref{eq33}) can be written as
 \begin{equation}
    \begin{aligned}
\left ( \omega _n^2-\omega ^2-i\frac{\omega \omega _n}{Q_n}  \right ) E_n(\omega )\int d^3x\mathbf{e} _n(\mathbf{x} )\cdot \mathbf{e} _n^*(\mathbf{x} )\\=\frac{g_{a\gamma \gamma }B_0\omega ^2}{\int d^3x\mathbf{e} _n(\mathbf{x} )\cdot \mathbf{e} _n^*}a(\omega )\int d^3x\vec{\mathbf{z}} \cdot \mathbf{e} _n^*(\mathbf{x} )~,
    \end{aligned}
 \end{equation}
which leads to
 \begin{equation}
    \begin{aligned}
 E_n(\omega )=\frac{g_{a\gamma \gamma }B_0}{\int d^3x\mathbf{e} _n(\mathbf{x} )\cdot \mathbf{e} _n^*(\mathbf{x} )}\frac{\omega ^2a(\omega )\int d^3x\vec{\mathbf{z}} \cdot \mathbf{e} _n^*(\mathbf{x} )}{\left ( \omega _n^2-\omega ^2-i\omega \omega _n/{Q_n}  \right )}~.
    \end{aligned}
 \end{equation}

Even if the axion field oscillates with a random phase, we may still assume that the long-time average of energy density repeats:
\begin{equation}
\begin{aligned}
\label{a^{2}}
\langle a^2\rangle={1\over 2T}\int_{-T}^Ta^2(t)dt=\int^{\infty}_{-\infty}|a(\omega)|^2{d\omega\over 2\pi}~.
\end{aligned}
\end{equation}
Because the time average of the electric field and magnetic field energy is the same in a cavity, the total energy $U$ in a cavity is
\begin{equation}
    \begin{aligned}
U =&\frac{g_{a\gamma \gamma }^2B_0^2|\int d^3x\vec{\mathbf{z}} \cdot \mathbf{e} _n^*(\mathbf{x} )|^{2}}{V\int d^3x\left |\mathbf{e} _n(\mathbf{x} )  \right | ^2 } \\ \times&V\int_{-\infty}^{\infty}\frac{d\omega}{2\pi}\frac{\omega^{4}|a(\omega)^{2}|}{\left | \omega _n^2-\omega ^2-i\omega \omega _n/{Q_n}   \right |^{2}}~.
    \end{aligned}
 \end{equation}
 For a high-quality factor \( Q \), the integral is predominantly concentrated around the resonant frequency \( \omega_{n} \), which yields:
\begin{equation}
\label{eqQ}
\begin{aligned}
\int_{-\infty}^{\infty}\frac{d\omega}{2\pi}\frac{\omega^{4}|a(\omega)^{2}|}{(\omega _n^2-\omega ^2)^{2}+\omega^{2}\omega _n^{2}/{Q_n^{2}}}\approx Q_{c}^{2}\left \langle a^{2} \right \rangle~.
\end{aligned}
\end{equation}
where $Q_c$ is the quality factor of the resonant cavity with detector loads. Inserting Eq.(\ref{EQ1}) to \( \left \langle a^{2} \right \rangle \) , because \( j \neq j' \) term cancel each other, one gets:
\begin{equation}
    \begin{aligned}
    \label{sum1}
\left \langle a^{2}\right \rangle=\frac{\rho_{DM}}{2T\cdot m^{2}}\cdot\sum_{j} \alpha_{j}^{2} f(v_{j})\Delta v\cdot\int_{-T}^{T}\cos^{2}(\omega t+\phi_{j})dt~.
    \end{aligned}
 \end{equation}
For a long time period, $\int \cos^{2}(\omega t + \phi_{j}) \, dt \approx T$, consequently:
\begin{equation}
\label{eqa}
    \begin{aligned}
\left \langle a^{2}\right \rangle\approx\frac{\rho_{DM}}{2m^{2}}\left \langle \alpha_{j}^{2}\right \rangle\int f(v)dv
\approx\frac{\rho_{DM}}{m^{2}}~,
    \end{aligned}
 \end{equation} where \( \left \langle \alpha^{2} \right \rangle = \int \alpha^{2} P(\alpha) \, d\alpha = 2 \).
$\alpha_j$ is a random variable (Rayleigh-distributed) sampled for each particle $j$ and is not time-dependent. The ensemble average is taken over $\alpha_j$ and $v_j$. The sum $\sum_j \alpha_j^2 f(v_j) \Delta v$ represents the ensemble average over the random phases and velocities of the dark matter particles. For large particle counts, $\sum_j \alpha_j^2 f(v_j) \Delta v \to \int \alpha^2_v f(v) dv = \bar{\alpha}^2 \int f(v) dv  $. There are differences between $\bar{\alpha}^2$ and $\langle \alpha^2 \rangle$ because the dark matter velocity distribution (e.g., truncated Maxwellian) is not exactly the same as the Rayleigh distribution. However, the difference is typically small. Note that the approximation in Eq.(\ref{eqa}) arises from the use of a special model, while the equality should be exact in physical context, $m^2\langle a^2\rangle=\rho_{DM}$ if dark matter is non-relativistic. 

Define a form factor:
\begin{equation}
    \begin{aligned}
C=\frac{\left | \int d^3x\vec{\mathbf{z}} \cdot \mathbf{e} _n(\mathbf{x} ) \right |^2 }{V\int d^3x\left |\mathbf{e} _n(\mathbf{x} )  \right | ^2 } ~,
    \end{aligned}
 \end{equation} 
and assuming the velocity distribution function is sharply peaked near the resonant point \( \int f(v) \, dv = 1 \), then the total energy stored in the cavity is
\begin{equation}
    \begin{aligned}
U=\frac{g_{a\gamma \gamma }^2B_0^2\rho _{DM}Q_{c}^{2}VC}{m^{2}}~,
     \end{aligned}
 \end{equation}
which leads to the power of axion-photon conversion:
 \begin{equation}
    \begin{aligned}
P_{axion}=\frac{U}{Q_{c}}m =\frac{g_{a\gamma \gamma }^2B_0^2\rho_{DM} VC}{m}Q_c~.
     \end{aligned}
 \end{equation}
The signal power is identical to what would be obtained for a coherent field. This arises because the random phases \(\phi_j\) in the non-coherent field cause cross-terms (\(j \neq j'\)) to cancel, leaving only the sum of squared amplitudes (diagonal terms). In addition, the ensemble average over \(\alpha_j\) and velocity distribution \(f(v)\) recovers the same mean energy density as a coherent field. Thus axion signal power hold for both coherent and non-coherent cases.

If the cavity narrower than dark matter linewidth, i.e., \(Q_c > Q_{DM}\), the cavity samples only a fraction \( \Delta \omega_c / \Delta \omega_{\text{DM}} = Q_{DM} / Q_c\) of dark matter power. In Eq. (\ref{eqQ}) \(a(\omega)^2\) is replaced by \(a(\omega)^2Q_{DM}/Q_c\), therefore, the conversion power is \cite{Kim:2020kfo}
\begin{equation}
    \begin{aligned}
P_{axion}=\frac{g_{a\gamma \gamma }^2B_0^2\rho_{DM} VC}{m}\mathrm{min}(Q_c,Q_{DM})~.
     \end{aligned}
 \end{equation}
 
\section{dark photon to photon transition in a resonant cavity}
\label{sec6}
Dark photons and Standard Model (SM) photons can be coupled through ultraviolet (UV) quantum processes, leading to a small kinetic mixing between them\cite{HOLDOM1986196}. In addition, dark photons can be produced in the early universe, for example, through the misalignment mechanism, and thus constituting a substantial portion of dark matter today. The dark matter dark photons are primarily dominated by their electric field component. Therefore, when a resonant cavity is present, the dark electric field can induce a weak electric current that generates photons within the cavity. The effective Lagrangian can be written as:
\begin{equation}
\begin{aligned}
\mathcal{L} =&-\frac{1}{4} \left({F} _{\mu\nu}{F} ^{\mu\nu}+{X} _{\mu\nu}{X} ^{\mu\nu}\right)+\frac{m^2}{2}{X} ^{\mu}{X} _{\mu}\\
&+m^2\chi{X} _{\mu}{A} ^{\mu} +J^{\mu} {A} _{\mu}~,
\end{aligned}
\end{equation}
where \( A_{\mu} \) and \( F_{\mu\nu} \) are the ordinary photon field and its field strength tensor, respectively; \( J^{\mu} \) is the ordinary current; \( X_{\mu\nu} \) and \( X_{\mu} \) are the dark photon tensor and dark photon field, respectively. The time-like polarization of dark photon \( X^0 \), is generally suppressed, and the dark matter dark photons are dominated by their electric field component. Thus, the equation of motion can be expressed as:
\begin{equation}
    \begin{aligned}
    \label{interaction term}
\partial_{\mu}\partial^{\mu}\vec{E} &=\chi m^2\vec{E'}~,
    \end{aligned}
\end{equation}
where $\chi$ is the kinetic mixing parameter between dark photon and photon\cite{DIENES1997104,Goodsell:2009xc}, and $\vec{E'}$ denotes the dark electric field. Using Eq.(\ref{EQ1}) $\vec{E'}$ can be written as:
\begin{equation}
    \begin{aligned}
    \label{E}
\vec {E'}&=-i\frac{\partial\vec{X}}{\partial t}\\
&\approx-i\sqrt{\rho_{DM}}\sum_{j}^{} \alpha _j\sqrt{f(v_j)\Delta v} \cdot \cos \left (\omega t+\phi _{j}    \right )\cdot \mathbf{\hat{n}} ~.
   \end{aligned}
 \end{equation}
An additional order-one amplitude factor could appear, depending on whether the polarization of the dark electric field is aligned or points in random directions. Denote $\omega=m(1+v_{j}^{2}/2)$. Let us expand the electric field $E$ into the cavity modes:
\begin{equation}
    \begin{aligned}
\vec{E} (\mathbf{x},t )=\sum_{k} E_k(t)\mathbf{e} _k(\mathbf{x} )~.
    \end{aligned}
 \end{equation}
When close to the cavity resonant frequency, the damping coefficient can be written as $\omega_k/Q_c$, so we have:
\begin{equation}
    \begin{aligned}
{d^2E _k(t)\over dt^2}+\frac{\omega _k}{Q_{c}} {dE_k(t)\over dt}+\omega _k^2E _k(t)=&\chi m^2 \frac{\int d^3x\mathbf{e} _k(\mathbf{x} )\cdot \vec {E}'}{\int d^3x\left | \mathbf{e} _k(\mathbf{x} ) \right | ^2}~.
    \end{aligned}
 \end{equation}
The cavity electric field spectrum is then:
\begin{equation}
    \begin{aligned}
E _{k}(\omega)=\frac{\chi m^2 {E}'(\omega)}{\omega_k^2- \omega^2 -i\omega_k \omega /Q_{c} } \frac{\int d^3x\mathbf{e} _k(\mathbf{x} )\cdot \hat n}{\int d^3x\left | \mathbf{e} _k(\mathbf{x} ) \right | ^2} ~.
    \end{aligned}
 \end{equation}
Assuming the average dark electric field energy density remains periodical, we then have:
\begin{equation}
\langle E'(t)^2\rangle={1\over 2T}\int_{-T}^TE'^2(t)dt=\int^{\infty}_{-\infty}|E(\omega)|^2{d\omega\over 2\pi}~.
\end{equation}
The energy stored in the cavity is
\begin{equation}
\begin{aligned}
U=&\int_{-\infty}^{\infty}\frac{d\omega}{2\pi}|E_{k}(\omega)|^{2}\int d^{3}x|\mathbf{e}_{k}(\mathbf{x})|^{2}\\
=&\chi^{2}m^{4}\cdot \frac{\left | \int d^3x\mathbf{e} _k(\mathbf{x} )\cdot   \mathbf{\hat{n}} \right |^2 }{\int d^3x\left | \mathbf{e} _k(\mathbf{x} ) \right | ^2}\\
&\cdot\int_{-\infty}^{\infty}\frac{d\omega}{2\pi}\frac{|E'(\omega)|^{2}}{(\omega^{2}_{k}-\omega^{2})^{2}+\omega_{k}^{2}\omega^{2}/Q_{c}^{2}}~.
    \end{aligned}
\end{equation}
For a high quality factor, the integral is predominantly concentrated around the resonant frequency $\omega_{k}$, so we have:
\begin{equation}
\begin{aligned}
U\approx\chi^{2}m^{4}\frac{Q_{c}^{2}}{\omega_{k}^{4}}\frac{\left | \int d^3x\mathbf{e} _k(\mathbf{x} )\cdot   \mathbf{\hat{n}} \right |^2 }{\int d^3x\left | \mathbf{e} _k(\mathbf{x} ) \right | ^2}\cdot\left \langle E'(t)^{2} \right \rangle~.
\end{aligned}
\end{equation}
Taking a similar calculation in the previous section, we obtain:
\begin{equation}
\begin{aligned}
\left \langle E'(t)^{2} \right \rangle=&\rho_{DM}\cdot\sum_{j}\alpha^{2}_{j}\ f(v_{j})\Delta v\cdot\frac{1}{2T}\int_{-T}^{T}\cos^{2}(\omega t+\phi_{j})dt\\
\approx&\rho_{DM}~.
\end{aligned}
\end{equation}
Let us define the form factor as:
\begin{equation}
\mathcal{G} =\frac{\left | \int d^3x\mathbf{e}_{k}(\mathbf{x} )\cdot   \mathbf{\hat{n}} \right |^2 }{V\int d^3x\left | \mathbf{e}_{k} (\mathbf{x} ) \right | ^2}~.
  \end{equation}
Thus, the total energy stored in the cavity is:
\begin{equation}
\begin{aligned}
U=&\chi^{2}m^{4}\rho_{DM}\frac{Q_{c}^{2}}{\omega_{k}^{4}}V\mathcal{G}\\
\approx&\chi^{2}\rho_{DM}Q_{c}^{2}V\mathcal{G}~.
\end{aligned}
\end{equation}
Finally, the signal power from dark photon-to-photon conversion is
\begin{equation}
\label{eq53}
    \begin{aligned}
P_{dp}=\frac{U}{Q_{c}}\omega _k=\chi ^2m\rho_{DM}V\mathcal{G}\mathrm{min}(Q_c,Q_{DM}) ~,
     \end{aligned}
 \end{equation}
where the $Q_c> Q_{DM}$ case has been incorporated, as discussed in the previous section.

\section{Scanning rate}
\label{sec70}
The signal-to-noise ratio (SNR) is given by \cite{Kim:2020kfo}:
\begin{equation}\label{SNR}
\mathrm{SNR} =\frac{P_{r}}{k_BT_{sys}\sqrt{\Delta\nu_{sign}/\Delta t}} ~,
\end{equation}
where $P_{r}$ is the signal power coupled to the receiver, $k_{B}$ is the Boltzmann constant and $T_{sys}$ is the equivalent system noise temperature. The denominator represents the noise of the system, which is described by the Dicke radiometer equation as fluctuations in the noise power over an
integration time $\Delta t$ within a frequency bandwidth $\Delta\nu_{sign}$. In current axion or dark photon experiments, the frequency bandwidth is typically set as $\Delta \nu_{sign}\approx \nu_0/10^6$ \cite{Dicke:1946glx}, where $\nu_0$ is the central frequency of the dark matter field.

The total power from the cavity can be divided into two components: (1) dissipation and (2) useful power received by the detector. The quality factors satisfy:
 \begin{equation}
 \label{QuanlityQ}
\begin{aligned}
\frac{1}{Q_c}=\frac{1}{Q_0}+\frac{1}{Q_r}~,
\end{aligned}
\end{equation}
where $Q_0$ represents the power loss due to dissipation, and $Q_r$ represents the power loss due to the receiver. A dimensionless cavity coupling parameter can be defined as $\beta=Q_0/Q_r$, and the signal power coupled to the receiver is given by:
\begin{equation}
\label{the signal power coupled to the receiver}
\begin{aligned}
P_{r}=\frac{\beta}{1+\beta}P_{DM}~.
\end{aligned}
\end{equation}
In the case that $Q_c<Q_{DM}$, combining Eq.(\ref{SNR}) and Eq.(\ref{the signal power coupled to the receiver}), we obtain:
\begin{equation}
\begin{aligned}
\frac{\Delta\nu_{sign}}{\Delta t}=\frac{\nu_0/Q_{DM}}{\Delta t}
=\frac{P_{axion/dp}^{2}}{k_{B}^{2}T_{sys}^{2}\mathrm{SNR}^{2}}\frac{\beta^{2}}{(1+\beta)^{2}}.
\end{aligned}
\end{equation}

Typical haloscope experiments use a copper cavity, where $Q_c\sim 10^{5}\ll Q_{DM}\sim 1/\delta v^2\sim 10^{6}$. Each frequency pin covers a frequency range $\Delta \nu_{cavi} =\nu _0/Q_{c}$. Depending on the particular scanning setups, the limit of scanning rate is then:
\begin{equation}
\label{scanning rate of axion}
\begin{aligned}
&R_{axion}= \frac{\Delta\nu_{cavi}}{\Delta t} = {\nu_0/Q_{DM}\over \Delta t}{Q_{DM}\over Q_c}\\
&=g_{a\gamma\gamma}^4 \frac{\rho_a^2}{m_a^2} \frac{1}{\mathrm{SNR}^2} \frac{B_0^4 V^2 C^2}{k_B^2 T_{sys}^2} \frac{\beta^2}{(1+\beta)^2} Q_{DM}Q_{c}~
\end{aligned}
\end{equation}
for axions, and
\begin{equation}
\label{scanning rate of dark photon}
\begin{aligned}
R_{dp}=\chi^4\rho_D^2 m_D^2\frac{1}{\mathrm{SNR}^2}\frac{V^2\mathcal{G}^2}{k_B^2T_{sys}^2}\frac{\beta^2}{(1+\beta)^2}Q_{DM}Q_c~
\end{aligned}
\end{equation}
for dark photons. In case $Q_c>Q_{DM}$, the signal bandwidth becomes the same as the cavity bandwidth: 
\begin{equation}
\begin{aligned}
\frac{\Delta\nu_{sign}}{\Delta t}=\frac{\nu_0/Q_{c}}{\Delta t}
=\frac{P_{DM}^{2}}{k_{B}^{2}T_{sys}^{2}\mathrm{SNR}^{2}}\frac{\beta^{2}}{(1+\beta)^{2}}.
\end{aligned}
\end{equation}
The scanning rate is then \cite{Kim:2020kfo}:
\begin{equation}
\label{scanning rate of axion}
\begin{aligned}
&R_{axion}=  {\nu_0/Q_{c}\over \Delta t}\\
&=g_{a\gamma\gamma}^4 \frac{\rho_a^2}{m_a^2} \frac{1}{\mathrm{SNR}^2} \frac{B_0^4 V^2 C^2}{k_B^2 T_{sys}^2} \frac{\beta^2}{(1+\beta)^2} Q_{DM}^2~
\end{aligned}
\end{equation}
for axions, and
\begin{equation}
\label{scanning rate of dark photon}
\begin{aligned}
R_{dp}=\chi^4\rho_D^2 m_D^2\frac{1}{\mathrm{SNR}^2}\frac{V^2\mathcal{G}^2}{k_B^2T_{sys}^2}\frac{\beta^2}{(1+\beta)^2}Q_{DM}^2~
\end{aligned}
\end{equation}
for dark photons. Thus, a higher cavity quality factor improves the scanning rate but saturates when $Q_c$ becomes larger then $Q_{DM}$. Therefore, a better understanding of the dark matter wave properties, such as its effective bandwidth, can optimize the experimental setups. It is important to emphasize that a non-coherent wave does not necessarily result in a larger bandwidth. "Non-coherent" here means the relative phases of the particles are not fixed, but their velocity distributions can still be small.    
\section{Conclusions}
\label{sec7}
Bosonic dark matter with sub-eV mass, such as axions, axion-like particles, and dark photons, is of great interest from both theoretical and experimental perspectives. These particles could constitute a substantial portion of dark matter today, with non-thermal production mechanisms that lead to a broad range of possible parameter spaces. Their extremely weak interaction with Standard Model particles makes cavity resonance enhancement essential to achieve the required sensitivity.

This paper demonstrates that the assumption of a coherent wave is not necessary for cavity resonance experiments, as long as an appropriate readout mechanism is employed. The resulting power in the cavity is the same for both coherent and non-coherent waves. From a classical perspective, this occurs because the random phases in the non-coherent field cause cross-terms to cancel, leaving only the sum of squared amplitudes. The ensemble average of the particles recovers the same mean energy density as in the coherent case. Therefore, the signal power holds for both coherent and non-coherent waves.

This concept becomes clearer from a quantum perspective \cite{Yang:2022uil,Zheng:2024kxn}, where the cavity's boost effects is similar to the Purcell effect, enhancing the density of states near the resonance frequency. Thus, transitions are boosted regardless of the coherence of the dark matter wave. Experimentally, this allows for longer integrate times without concern for the decoherence of dark matter waves. 

It is important to note that bandwidth and coherence should be considered as independent features in cavity experiments. The bandwidth of dark matter waves is crucial to these experiments, while coherence is not. Naturally, a non-coherent dark matter wave will produce a non-coherent microwave signal in the cavity, meaning the readout mechanism may need to be adapted (e.g. using a single-photon detector). This makes cavity experiments applicable to a broader range of scenarios, and the integration time could be extended. Combining this understanding with quantum theory and quantum measurement techniques presents significant potential for future development in cavity experiments.

\bibliography{dp}
\end{document}